\documentclass[letterpaper,11pt]{article}
\usepackage{amssymb}
\usepackage{amsmath}
\usepackage{amstext}
\usepackage{graphicx,epsfig}
\usepackage{epsfig}
\usepackage{verbatim} 
\usepackage{fancybox}
\usepackage{color}
\usepackage{ulem}
\usepackage{enumitem}
\usepackage{subfigure}
\usepackage{bbm}
\usepackage{parskip}
\usepackage{dsfont}
\usepackage[numbers,sort&compress]{natbib}
\usepackage[body={6.5in, 9in}, right=1in, top=1in]{geometry}

\newcommand{\be}{\begin{equation}}
\newcommand{\ee}{\end{equation}}
\newcommand{\bea}{\begin{eqnarray}}
\newcommand{\eea}{\end{eqnarray}}
\newcommand{\beas}{\begin{eqnarray*}}
\newcommand{\eeas}{\end{eqnarray*}}

\def\({\left(}
\def\){\right)}

\def\gsim{ \lower .75ex \hbox{$\sim$} \llap{\raise .27ex \hbox{$>$}} }
\def\lsim{ \lower .75ex \hbox{$\sim$} \llap{\raise .27ex \hbox{$<$}} }

\usepackage{xcolor,colortbl}

\begin{document}
\def\thefootnote{\fnsymbol{footnote}}

\begin{center}
\Large{\textbf{Another Path for the Emergence of Modified Galactic Dynamics from Dark Matter Superfluidity}} \\[0.5cm]
 
\large{Justin Khoury}
\\[0.5cm]

\small{
\textit{Center for Particle Cosmology, Department of Physics and Astronomy, \\ University of Pennsylvania, Philadelphia, PA 19104}}

\vspace{.2cm}

\end{center}

\vspace{.2cm}

\hrule \vspace{0.2cm}
\centerline{\small{\bf Abstract}}
{\small In recent work we proposed a novel theory of dark matter (DM) superfluidity that matches the successes of the $\Lambda$CDM model on cosmological scales while simultaneously reproducing MOdified Newtonian Dynamics (MOND) phenomenology on galactic scales. The agents responsible for mediating the MONDian force law are superfluid phonons that couple
to ordinary (baryonic) matter. In this paper we propose an alternative way for the MOND phenomenon to emerge from DM superfluidity. The central idea is to use higher-gradient corrections in the superfluid effective theory. These next-to-leading order terms involve gradients of the gravitational potential, and therefore effectively modify the gravitational force law. In the process we discover a novel mechanism for generating the non-relativistic MOND action, starting from a theory that is fully analytic in all field variables. The idea, inspired by the symmetron mechanism, uses the spontaneous breaking of a discrete symmetry. For large acceleration, the symmetry is unbroken and the action reduces to Einstein gravity. For small acceleration, the symmetry is spontaneously broken and the action reduces to MONDian gravity. 
Cosmologically, however, the universe is always in the Einstein-gravity, symmetry-restoring phase. The expansion history and linear growth of density perturbations are therefore indistinguishable from $\Lambda$CDM cosmology.}   
\vspace{0.3cm}
\noindent
\hrule
\def\thefootnote{\arabic{footnote}}
\setcounter{footnote}{0}

\section{Introduction}

Much work has been done~\cite{Blanchet:2006yt,Blanchet:2008fj,Zhao:2008rq,Bruneton:2008fk,Li:2009zzh,Ho:2010ca,Ho:2011xc,Ho:2012ar,Khoury:2014tka} to reconcile the phenomenological success of Cold Dark Matter (CDM) on cosmological scales with the empirical success of MOdified Newtonian Dynamics (MOND)~\cite{Milgrom:1983ca,Milgrom:1983pn,Milgrom:1983zz,Bekenstein:1984tv} on galactic scales. Briefly, the MOND force law reduces to Newtonian gravity at large acceleration but departs from Newton at low acceleration, with critical acceleration $a_0$:
\be
a =\left\{\begin{array}{cl} 
a_{\rm N} \qquad ~~~~a_{\rm N} \gg a_0\\[.5cm]
\sqrt{a_{\rm N}a_0} \qquad a_{\rm N} \ll a_0 \,.
\end{array}\right.
\label{MONDlaw}
\ee
This empirical law has been remarkably successful at explaining a wide range of galactic phenomena~\cite{Famaey:2011kh}, with the best-fit value for $a_0$ intriguingly of order the present Hubble parameter:
\be
a_0 \simeq \frac{1}{6}H_0 \simeq  1.2\times 10^{-8}~{\rm cm}/{\rm s}^2 = 2.7 \times 10^{-34}~{\rm eV}\,. 
\label{a0}
\ee
In particular, the observed Baryonic Tully-Fisher Relation~\cite{Tully:1977fu,Freeman1999,McGaugh:2000sr,McGaugh:2005qe,McGaugh:2011ac}, $M_{\rm b} \sim v^4_{\rm c}$, relating the total baryonic mass to the asymptotic circular velocity, immediately follows from this force law.

In a series of recent papers~\cite{Berezhiani:2015pia,Berezhiani:2015bqa,Khoury:2015pea} the present author together with Berezhiani (hereafter BK) proposed a novel framework that unifies DM and MOND through the physics of superfluidity. The two seemingly unrelated phenomena in fact have a common origin, as different phases of a single underlying substance.\footnote{The possible connection between MOND and superfluidity was mentioned already years ago by Milgrom in~\cite{Milgrom:2001gz}, though of course not in the context of DM.}
Traditionally thought of as a modification to General Relativity, MOND requires new degrees of freedom (in particular a scalar field) to mediate an
additional force beyond gravity. The MOND scalar field is classical for all situations of interest. The key observation of BK
is that superfluidity naturally offers a coherent, classical scalar field --- the phase of the condensate wavefunction, whose excitations are phonons. Therefore, if DM is a superfluid (of a suitable kind)
with a coherence length of order galactic scale, then its phonon excitations can generate the desired MOND phenomenon, without the need for additional degrees of freedom. 

The possibility of DM forming a Bose-Einstein condensate (BEC) in galaxies has of course been studied before, {\it e.g.}~\cite{Silverman:2002qx,vortex2,Boehmer:2007um,Sin:1992bg,Ji:1994xh,Hu:2000ke,Goodman:2000tg,Peebles:2000yy,Arbey:2003sj,Lee:2008ux,Lee:2008jp,Harko:2011xw,Dwornik:2013fra,Guzman:2013rua,Harko:2014vya}, but these earlier papers focused primarily on the condensate profile to explain galactic rotation curves, with phonons being irrelevant. The idea that phonons can play a key role by altering the motion of ordinary matter and yielding an effective MOND law is to our knowledge new. 

As discussed in BK, superfluidity will arise if DM particles: $i)$ are sufficiently light ($m \;\lsim\; 2~{\rm eV}$) such that 
their de Broglie wavelength $\lambda_{\rm dB} \sim \frac{1}{mv}$ overlaps in galaxies; $ii)$ interact sufficiently strongly ($\frac{\sigma}{m}\; \gsim\; 0.1~{\rm cm}^2/{\rm g}$) to
establish thermal equilibrium. Therefore, DM consists of strongly interacting axion-like particles. The critical temperature is then found to be $\sim {\rm mK}$, which intriguingly is comparable to cold atom gases studied in the laboratory. In particular, superfluidity (and along with it MOND) only occurs in sufficiently low-mass halos, where the DM temperature (set by the virial velocity) is below critical. With $m \sim {\rm eV}$, the threshold is around $10^{12}-10^{13}~M_\odot$. In that case massive galaxy clusters (with $M \;\gsim\; 10^{14}~M_\odot$) are above critical temperature, and DM is in the normal phase. 
Thus the framework successfully distinguishes between galaxies (where MOND is successful) and galaxy clusters (where MOND is not). 

In this paper we present an alternative mechanism for realizing MOND within DM superfluidity. Instead of relying on the mediation of a ``fifth force" by phonons, as in BK, we propose 
to use higher-gradient corrections in the effective theory of superfluidity to generate the MOND phenomenon. These next-to-leading order terms involve gradients of the gravitational potential,
$\vec{\nabla}\Phi$, and therefore effectively modify gravity. The required form of these corrections is admittedly special, but we will argue that the effective theory is under control both in the Newtonian and
deeply-MONDian asymptotic regimes. The proposal has much in common with BK --- in particular everything from the previous paragraph carries over --- but there are
also important differences:

\begin{itemize}

\item The most important difference is that the modified force law~\eqref{MONDlaw}, being realized through the gravitational potential, is now {\it universal}. It applies not only to baryonic matter
but also to DM. This modifies the density profile of DM superfluid halos, compared to those of BK. 

\item Phonons are no longer responsible for MOND. This has two implications. First, we have more freedom in specifying the superfluid equation of state and are not confined to
the particular form studied in BK. In our analysis we will consider a general polytropic equation of state $P \sim \rho^\alpha$. Second, phonons need not couple directly to ordinary 
matter, and as a result enjoy an exact shift symmetry. 

\item In BK, the stability of the MOND-like solution hinges on finite-temperature corrections. These corrections are expected to contribute at some level since DM in galactic halos
has non-zero temperature, as mentioned earlier, and it is reasonable to postulate that they can stabilize perturbations. That said, it would be more satisfactory if the zero-temperature
theory were stable all by itself. We will see that this is the case in the present scenario. 

\end{itemize}

In traditional formulations of MOND, the interpolating function necessary to realize~\eqref{MONDlaw} is usually
non-analytic in $(\vec{\nabla}\Phi)^2$ or other field variables. As it does not obviously lend itself to the rules
of effective field theory. In this paper, we will present a novel mechanism for generating the non-relativistic MOND action,
starting from a theory that is fully analytic in all field variables. This mechanism will of course be described in the
superfluid context, but is in fact completely independent and applies more generally. 

The idea is inspired by the symmetron mechanism~\cite{Hinterbichler:2010es,Hinterbichler:2011ca,Olive:2007aj,Pietroni:2005pv}.
It uses a scalar field $\chi$ whose action is invariant under a discrete $\mathbb{Z}_2$ symmetry $\chi\rightarrow -\chi$. Whether
this symmetry is spontaneously broken or not depends on the sign of the mass squared around $\chi = 0$. The scalar field
couples non-minimally to gravity and to the DM superfluid, and as a result its mass is controlled by $\frac{|\vec{\nabla}\Phi|}{a_0}$.
For $|\vec{\nabla}\Phi| > 3a_0$, the mass squared is positive, and the symmetry is unbroken. In this case, the
action reduces to Newtonian gravity. For $|\vec{\nabla}\Phi| < 3a_0$, the mass squared is negative, and the symmetry is
spontaneously broken. The action then reduces to MOND. Perturbations around either vacuum are stable and subluminal. 

Aside from having an analytic action, the symmetron-inspired mechanism nicely circumvents an important challenge
for any hybrid scenario that contains both MOND and DM. Namely, while one wants MOND to dominate the
dynamics in galaxies, on cosmological scales the evolution should be primarily dictated by DM evolving according to
standard gravity in order to reproduce the matter power spectrum and other linear scale observables. In the BK scenario, for instance,
this requires the critical acceleration $a_0$ and the phonon-matter coupling parameter to depend on
temperature~\cite{Berezhiani:2015pia,Berezhiani:2015bqa}.  In our case we will find that on a cosmological background
the mass squared for $\chi$ is positive definite at all times. Therefore, cosmologically the universe is always in the Einstein-gravity,
symmetry-restoring phase. Consequently the expansion history and linear growth of density perturbations are indistinguishable from $\Lambda$CDM cosmology.

The paper is organized as follows. After brief reviewing the BK scenario in Sec.~\ref{BKreview}, we describe in Sec.~\ref{newscenario} 
the present approach to MOND using next-to-leading order corrections in the superfluid effective theory. In particular, we present the
symmetron-inspired mechanism to achieve MOND as a phase of spontaneous symmetry breaking. Section~\ref{paramcons} discusses
various phenomenological constraints on the parameters of the theory. In Sec.~\ref{profile} we derive the DM superfluid density profile for
a static, spherically-symmetric halo, assuming hydrostatic static equilibrium and MONDian gravity. We use the result to calculate the halo radius
and central density as a function of the parameters of the theory. In Sec.~\ref{lensing} we present a relativistic version of the theory, which enforces
the equality of gravitational potentials, required for dynamical and lensing mass estimates to coincide. Section~\ref{cosmo} is devoted to cosmology.
In particular, we will show that on linear scales the symmetry is always restored and gravity is Einsteinian. In other words, the MOND phenomenon only occurs
in regions where spatial gradients  of the metric dominate over its time derivatives, {\it i.e.}, in non-linear structures. We close with a few concluding
remarks in Sec.~\ref{conclu}.

\section{MOND from phonons}
\label{BKreview}

We briefly review the BK framework of superfluidity~\cite{Berezhiani:2015pia,Berezhiani:2015bqa,Khoury:2015pea}. In the language of field theory, an (abelian) superfluid is described by the theory of 
a spontaneously broken global $U(1)$ symmetry, in a state of finite charge density.  The relevant degree
of freedom at low energy is the Goldstone boson for the broken symmetry, namely the phonon field $\theta$. The $U(1)$ symmetry
acts non-linearly on $\theta$ as a shift symmetry, $\theta \rightarrow \theta + c$. In the non-relativistic regime and in the absence of external potentials,
the theory should be Galilean invariant. According to the rules of effective field theory, we are instructed to write down all possible operators consistent with these symmetries. 
We will be interested in the case where there is a gravitational potential $\Phi$. 

At leading order~(LO) in the derivative expansion, the relevant building block is the kinetic operator 
\be
X =  \dot{\theta}  - m \Phi - \frac{(\vec{\nabla}\theta)^2}{2m}\,.
\label{X}
\ee
The most general LO action is an arbitrary function of this quantity~\cite{Greiter:1989qb,Son:2002zn}:
\be
{\cal L}_{\rm LO} = P(X)\,.
\label{PXgen}
\ee
At finite chemical potential, $\theta = \mu t$, this action defines the grand canonical equation of state $P(\mu)$ of the superfluid. 
A straightforward calculation of the stress energy tensor reveals that the energy density is
\be
\rho = m P_{,X}\,,
\label{rhoP}
\ee
while the pressure is $P$. In other words, the type of superfluid is uniquely encoded in the choice of $P(X)$. 
Perturbations $\varphi = \theta - \mu t$ about this state describe phonon excitations. 

In BK we conjectured that DM phonons are described by the non-relativistic MOND
scalar action,\footnote{The square-root form is necessary to ensure that the action is well-defined for time-like field profiles, and that the Hamiltonian
is bounded from below~\cite{Bruneton:2007si}.}
\be
P(X) = \frac{2\Lambda(2m)^{3/2}}{3} X\sqrt{|X|}\,,
\label{PMOND}
\ee
corresponding to $P \sim \mu^{3/2}$. Using the thermodynamic relation~\eqref{rhoP}, this implies a polytropic equation of state $P\sim \rho^3$.

To mediate a MONDian force between ordinary matter, phonons must couple to baryons through 
\be
{\cal L}_{\rm int} \sim \frac{\Lambda}{M_{\rm Pl}} \theta \rho_{\rm b}\,,
\label{Lintintro}
\ee
which softly breaks the shift symmetry for $\theta$. With this action, the phonon-mediated force and the usual Newtonian gravitational force together
give an effective MOND force law, with the scale $\Lambda$ related to the critical acceleration via $\Lambda \sim \sqrt{a_0M_{\rm Pl}} \sim {\rm meV}$.  
Unlike ``pure" MOND, however, the DM halo itself contributes to the Newtonian component of the acceleration. This contribution is negligible on
distances probed by galactic rotation curves, but becomes comparable to the MOND component at distances of order the size of the superfluid core. 

The superfluid interpretation has a number of advantages over other formulations of MOND. For starters it is more economical. There is no need to postulate
additional degrees of freedom to modify gravity --- the coherent phonon scalar field is enough. Secondly, the non-analytic nature of the kinetic term~\eqref{PMOND}
is more palatable, as it is intrinsically tied to the superfluid equation of state. The MONDian action~\eqref{PMOND} corresponds to $P\sim \rho^3$, as mentioned earlier,
which is analytic. In fact there is a well-known example of a theory with fractional power in cold atom systems --- the Unitary Fermi Gas (UFG)~\cite{UFGreview,Giorgini:2008zz,Braaten:2004rn},
describing fermionic atoms at unitary. The UFG superfluid action is fixed by non-relativistic scale invariance to the non-analytic form ${\cal L}_{\rm UFG}(X) \sim X^{5/2}$~\cite{Son:2005rv}. 

While offering a tantalizing reconciliation of DM and MOND, the explicit realization~\eqref{PMOND} has a few features that could be improved upon:

\begin{itemize}

\item For a static, spherically-symmetric source, the scalar equation has two branches of solutions, depending on the sign of $X$. The $X< 0$ branch 
gives rise to a MONDian regime at low acceleration, while the $X>0$ branch does not. However, perturbations around the MONDian branch have wrong-sign kinetic term, signaling a ghost instability. As
shown in BK, this instability can be cured by finite-temperature corrections. These corrections are expected to contribute at some level since
DM particle in galactic halos have non-zero velocity and hence non-zero temperature. But it would be nice if the zero-temperature theory were by itself stable.

\item The baryon coupling~\eqref{Lintintro}, while technically natural from an effective field theory point of view, picks out a preferred phase of the wavefunction.
This seems unphysical. One possibility is that the coupling involves a {\it difference} of phase, which is a physical quantity, say between the local phase and the cosmological phase.
Another possibility is that the shift symmetry is broken to a discrete subgroup through a $\cos\theta\rho_{\rm b}$ operator~\cite{Villain,cosinecoupling,Randy}.
Expanding around the state at finite chemical potential $\theta = \mu t$, such a term would give~\eqref{Lintintro} to leading order, albeit with an oscillatory prefactor.

\end{itemize}
The goal of this paper is to offer an alternative way of realizing MOND in the superfluid context, which avoids all aforementioned drawbacks. 

\section{New approach: MOND from higher-derivative corrections}
\label{newscenario}

The aim of this work is to present an alternative mechanism, still based on DM superfluidity, for generating the MOND phenomenon. 
The central idea is to use higher-gradient corrections in the superfluid effective theory to obtain a MOND-like action directly in terms of the gravitational potential. 
In particular, phonons no longer couple directly to baryons and therefore are no longer responsible for mediating the MOND force.
The operator~\eqref{Lintintro} is absent, and the shift symmetry is now exact. In fact, in static situations and at finite chemical potential,
we will see that $\theta = \mu t$ is a solution, {\it i.e.}, phonon excitations can be consistently set to zero. 

The theory we have in mind consists of two terms:
\be
{\cal L} = {\cal L}_{\rm LO} + {\cal L}_{{\rm NLO}/{\rm grav}}\,.
\ee
We discuss each term below.

\subsection{Leading-order superfluid action}
\label{LO}

The first term, ${\cal L}_{\rm LO}$, describes the leading-order superfluid action~\eqref{PXgen}. 
Since phonons do not directly mediate the MOND force, there is much freedom in specifying the superfluid equation of state; 
we are no longer confined to the particular form $P(X)\sim X\sqrt{|X|}$. For simplicity, we will focus on power-law forms
\be
{\cal L}_{\rm LO} = \frac{\Lambda^4}{n} \left(\frac{X}{m}\right)^{n}\,.
\label{Pn}
\ee
(As a side bonus of this approach, we will always have $X>0$, hence square root forms like~\eqref{PMOND} are no longer necessary.)
Using~\eqref{rhoP} the energy density is
\be
\rho = \Lambda^4 \left(\frac{X}{m}\right)^{n-1}\,.
\label{rhosmu}
\ee
The equation of state is 
\be
P = \frac{\Lambda^4}{n} \left(\frac{\rho}{\Lambda^4}\right)^{\frac{n}{n-1}}\,.
\label{eos}
\ee
The adiabatic sound speed is given by
\be
c_s^2 = \frac{{\rm d}P}{{\rm d}\rho} = \frac{1}{n-1} \frac{X}{m}\,.
\label{cs}
\ee
Note that we need $n > 1$ in order for $c_s^2 > 0$. 

The action~\eqref{Pn} describes a theory of interacting phonons. Expanding in perturbations $\varphi = \theta - \mu t$ at finite chemical potential, 
we obtain
\be
{\cal L}_{\rm LO} = \frac{\Lambda^4\mu^{n-2}}{2m^n}\left(\dot{\varphi}^2 - c_s^2 (\vec{\nabla}\varphi)^2\right) + \sum_{k = 3}^\infty C_k \frac{\Lambda^4\mu^{n-k}}{m^n} (\partial \varphi)^k\,,
\ee
where $\partial$ stands for either $\partial_t$ or $c_s \vec{\nabla}$, and the $C_k$'s are order unity coefficients. After canonically-normalizing the kinetic term via
$\varphi_c \sim \sqrt{\frac{\Lambda^4\mu^{n-2}}{m^n}}\varphi$, we identify the strong coupling scale --- the scale at which perturbative unitarity breaks down --- as
the scale suppressing higher-dimensional operators:
\be
\Lambda_{\rm strong} \sim \Lambda \left(\frac{\mu}{m}\right)^{n/4} \sim \Lambda \, c_s^{n/2}\,.
\label{Lamstrong}
\ee

Two values of $n$ are of particular interest:

\begin{itemize}

\item $n = 2$: This equation of state, $P \sim \rho^2$, corresponds to a textbook BEC~\cite{Dalfovo:1999zz}. This was the case considered
in earlier studies of BEC DM, {\it e.g.},~\cite{Boehmer:2007um}. It arises in the weakly-coupled regime of bosons interacting through 2-body $s$-wave scattering.
In terms of the scattering length $a$, the equation of state $P = \frac{2\pi a}{m^3}\rho^2$ translates in our notation to $\Lambda^4 = \frac{m^3}{4\pi a}$.
The perturbative ``UV completion" with linearly-realized global $U(1)$ symmetry is a complex scalar field with $|\Psi|^4$ interactions.

\item $n = 5/2$: As mentioned earlier, this is the case relevant for the UFG~\cite{Son:2005rv}. It describes a gas of fermions
with 2-body interactions tuned to infinite scattering length, $a\rightarrow \infty$. The equation of state is $P\sim  \rho^{5/3}$.

\end{itemize}
In Sec.~\ref{profile} we will derive the spherically-symmetric halo density profile for each of these two cases. 

\subsection{Next-to-leading order and gravitational action}

At next-to-leading order (NLO), the action receives various corrections from higher-derivative operators,
such as $(\vec{\nabla} X)^2$ and $\vec{\nabla}^2\theta$, as well as possible non-minimal couplings to gravity~\cite{Son:2005rv}.
Our idea is to exploit this fact to generate a MOND-like action for $\Phi$. In particular, consider the operator $(\vec{\nabla} X)^2$. 
Working at finite chemical potential, $\theta = \mu t$ and ignoring phonon excitations, we have $X = \mu - m\Phi$, and therefore
\be
(\vec{\nabla} X)^2 \rightarrow m^2 (\vec{\nabla} \Phi)^2 \,.
\label{XPhi}
\ee
(Meanwhile, in the same limit $\vec{\nabla}^2\theta \rightarrow 0$.) Thus NLO corrections involving $(\vec{\nabla} X)^2$ effectively modify
the kinetic term for gravity.

To achieve MOND we conjecture that NLO superfluid terms mix with gravity as follows 
\be
{\cal L}_{{\rm NLO}/{\rm grav}} = -M_{\rm Pl}^2(\vec{\nabla}\Phi)^2 \, f((\vec{\nabla} X)^2)\,,
\ee
(We focus for the moment on the non-relativistic limit for gravity, and will come back to its relativistic generalization in Sec.~\ref{lensing}.) Via~\eqref{XPhi},
$f$ becomes a function of $(\vec{\nabla} \Phi)^2$ and plays the role of the interpolating function in MOND. Specifically, it should satisfy 
\be 
f((\vec{\nabla} X)^2) \simeq \left\{\begin{array}{cl}
1 \hspace{20pt}&\text{for}\hspace{10pt} \frac{|\vec{\nabla}X|}{m} \gg a_0\,, \\ \\
\frac{2}{3}\frac{\sqrt{(\vec{\nabla}X)^2}}{ma_0} \hspace{20pt}&\text{for}\hspace{10pt} \frac{|\vec{\nabla}X|}{m} \ll a_0 \,.
\end{array}\right.
\label{flimits}
\ee
A straightforward possibility, often assumed in the MOND literature~\cite{Famaey:2011kh}, is to postulate a function with these asymptotic behaviors. Unfortunately such a function
is inevitably non-analytic in $(\vec{\nabla} \Phi)^2$ (or $(\vec{\nabla} X)^2$ in our case), and as such does not obviously lend itself to the usual rules of effective field theory.

Here we propose an alternative path to~\eqref{flimits}, which to our knowledge is new, starting from an action that is analytic.
The idea is loosely based on the ``symmetron" mechanism~\cite{Hinterbichler:2010es,Hinterbichler:2011ca,Olive:2007aj,Pietroni:2005pv} proposed 
in a different context. Consider a dimensionless scalar field $\chi$ coupled non-minimally to gravity and to $(\vec{\nabla} X)^2$:
\be
{\cal L}_{{\rm NLO}/{\rm grav}} =- \frac{1}{2} Z^2 (\partial \chi)^2  -M_{\rm Pl}^2(\vec{\nabla}\Phi)^2 \left(\frac{1}{1 + \chi^2} + \frac{(\vec{\nabla} X)^2}{9m^2a_0^2}\chi^2\right)\,.
\label{NLOX}
\ee
The coefficient $Z$ of the kinetic term, which has dimension of mass, will be constrained by phenomenology in Sec.~\ref{paramcons}.
With the replacement~\eqref{XPhi}, $\chi$ is thus governed by an effective potential that depends on $(\vec{\nabla} \Phi)^2$:
\be
V(\chi) = M_{\rm Pl}^2(\vec{\nabla}\Phi)^2 \left(\frac{1}{1 + \chi^2} + \frac{(\vec{\nabla} \Phi)^2}{9a_0^2}\chi^2\right)\,.
\label{Veff18}
\ee

The action~\eqref{NLOX} is invariant under the $\mathbb{Z}_2$ symmetry $\chi\rightarrow -\chi$. Whether this symmetry 
is spontaneously broken or not depends on the sign of the mass squared around $\chi = 0$:
\be
m_\chi^2 = \left.\frac{{\rm d}^2V}{{\rm d}\chi^2}\right\vert_{\chi = 0} = 2M_{\rm Pl}^2(\vec{\nabla}\Phi)^2\left(-1 +  \frac{(\vec{\nabla} \Phi)^2}{9a_0^2}\right)\,.
\label{meffstatic}
\ee
(The {\it physical} mass after canonical normalization is $m_{\rm phys} = \frac{m_\chi}{Z}$.) Therefore, whether there is symmetry 
breaking or not depends on the magnitude of $|\vec{\nabla} \Phi|$ compared to $a_0$, with the phase transition 
occurring at the critical value $|\vec{\nabla} \Phi_{\rm c} | = 3 a_0$. This is closely analogous to the symmetron 
mechanism~\cite{Hinterbichler:2010es,Hinterbichler:2011ca,Olive:2007aj,Pietroni:2005pv}, except that symmetry
breaking is dictated here by $|\vec{\nabla} \Phi|$ instead of the matter density. 

In the Newtonian regime, $|\vec{\nabla} \Phi| > 3a_0$, the mass-squared is positive, and the potential is minimized at the symmetry-restoring point:
$\chi = 0$. In this unbroken phase, the action~\eqref{NLOX} reduces to standard Newtonian gravity:
\be
{\cal L}_{{\rm NLO}/{\rm grav}} \simeq - M_{\rm Pl}^2 (\vec{\nabla} \Phi)^2 \qquad ({\rm symmetry~restoring})\,.
\ee

In the MONDian regime, $|\vec{\nabla} \Phi| < 3a_0$, on the other hand, the mass term is negative, and $\chi$ acquires a vacuum expectation value (VEV):
\be
\chi = \pm  \sqrt{\frac{3a_0}{|\vec{\nabla} \Phi |}-1}\,.
\label{broken}
\ee
In this phase the $\mathbb{Z}_2$ symmetry is spontaneously broken. Substituting~\eqref{broken} into the action~\eqref{NLOX} (via~\eqref{XPhi}), and ignoring
higher-gradient terms arising from $(\partial\chi)^2$, we find
\be
{\cal L}_{{\rm NLO}/{\rm grav}} \simeq-\frac{2M_{\rm Pl}^2}{3a_0}\left((\vec{\nabla} \Phi)^2\right)^{3/2} + \frac{M_{\rm Pl}^2}{9}\frac{(\vec{\nabla} \Phi)^4}{a_0^2} \qquad ({\rm symmetry~breaking})\,.
\label{actiondeepMOND}
\ee
The first term is recognized as the MOND action, while the second term is subdominant in the deep MOND regime $|\vec{\nabla} \Phi| \ll a_0$.
Note that the non-analytic nature of the MOND action emerges from an original action that is completely analytic in $\chi$ and $(\vec{\nabla} \Phi)^2$.

\subsection{Summary}

Including the coupling of baryons to gravity, our complete non-relativistic action is given by
\be
{\cal L}  =  - \frac{1}{2} Z^2 (\partial \chi)^2  -M_{\rm Pl}^2(\vec{\nabla}\Phi)^2 \left(\frac{1}{1 + \chi^2} + \frac{(\vec{\nabla} X)^2}{9m^2a_0^2}\chi^2\right) + \frac{\Lambda^4}{n} \left(\frac{X}{m}\right)^n  - \Phi\rho_{\rm b}\,.
\label{Ltot}
\ee
At finite chemical potential, $\theta = \mu t$ ignoring phonon excitations, we can make the replacement $(\vec{\nabla} X)^2\rightarrow m^2 (\vec{\nabla} \Phi)^2$. Furthermore, the LO term can be expanded to linear order in $\Phi$. The action becomes (up to an irrelevant constant)
\be
{\cal L} = - \frac{1}{2} Z^2 (\partial \chi)^2  -M_{\rm Pl}^2(\vec{\nabla}\Phi)^2 \left(\frac{1}{1 + \chi^2} + \frac{(\vec{\nabla} \Phi)^2}{9a_0^2}\chi^2\right)  - \left(\rho_{\rm s}  + \rho_{\rm b}\right)\Phi\,,
\label{Ltotmu}
\ee
where $\rho_{\rm s} \equiv  \Lambda^4 \left(\frac{\mu}{m}\right)^{n-1}$ is the superfluid mass density~\eqref{rhosmu}.

In the Newtonian, symmetry-restoring phase ($|\vec{\nabla} \Phi| \;\gsim\; 3a_0$), we have $\chi = 0$, and the action reduces to
\be
{\cal L}_{\rm Newtonian} \simeq  - M_{\rm Pl}^2 (\vec{\nabla} \Phi)^2 - \left(\rho_{\rm s} + \rho_{\rm b}\right)\Phi \,.
\ee
This describes Newtonian gravity sourced by the DM superfluid and baryonic components, with standard Poisson equation:
\be
\vec{\nabla}^2\Phi = \frac{\rho_{\rm s} + \rho_{\rm b} }{2M_{\rm Pl}^2}\,.
\ee
On the other hand, in the deeply MONDian, symmetry-breaking phase ($|\vec{\nabla} \Phi| \ll 3a_0$), $\chi$ acquires the VEV~\eqref{broken}, and the effective action to leading order in gradients is
\be
{\cal L}_{\rm MOND} \simeq  -\frac{2M_{\rm Pl}^2}{3a_0}\left((\vec{\nabla} \Phi)^2\right)^{3/2}  -  \left(\rho_{\rm s} + \rho_{\rm b}\right)\Phi  \,.
\ee
This describes MONDian gravity coupled to both baryonic {\it and} DM components:
\be
\vec{\nabla}\cdot \left(\frac{|\vec{\nabla}\Phi|}{a_0}\,\vec{\nabla}\Phi\right) = \frac{\rho_{\rm s} + \rho_{\rm b} }{2M_{\rm Pl}^2}\,.
\ee

\section{Parameter Constraints}
\label{paramcons}

Other than the Planck mass and the critical acceleration $a_0$, our theory has 3 parameters: the DM particle mass $m$, the wavefunction renormalization $Z$ for $\chi$,
and the scale $\Lambda$ of the superfluid Lagrangian. In this Section we discuss various constraints on these parameters. 

\subsection{DM particle mass}

\begin{figure}[t]
\centering
\includegraphics[width=4in]{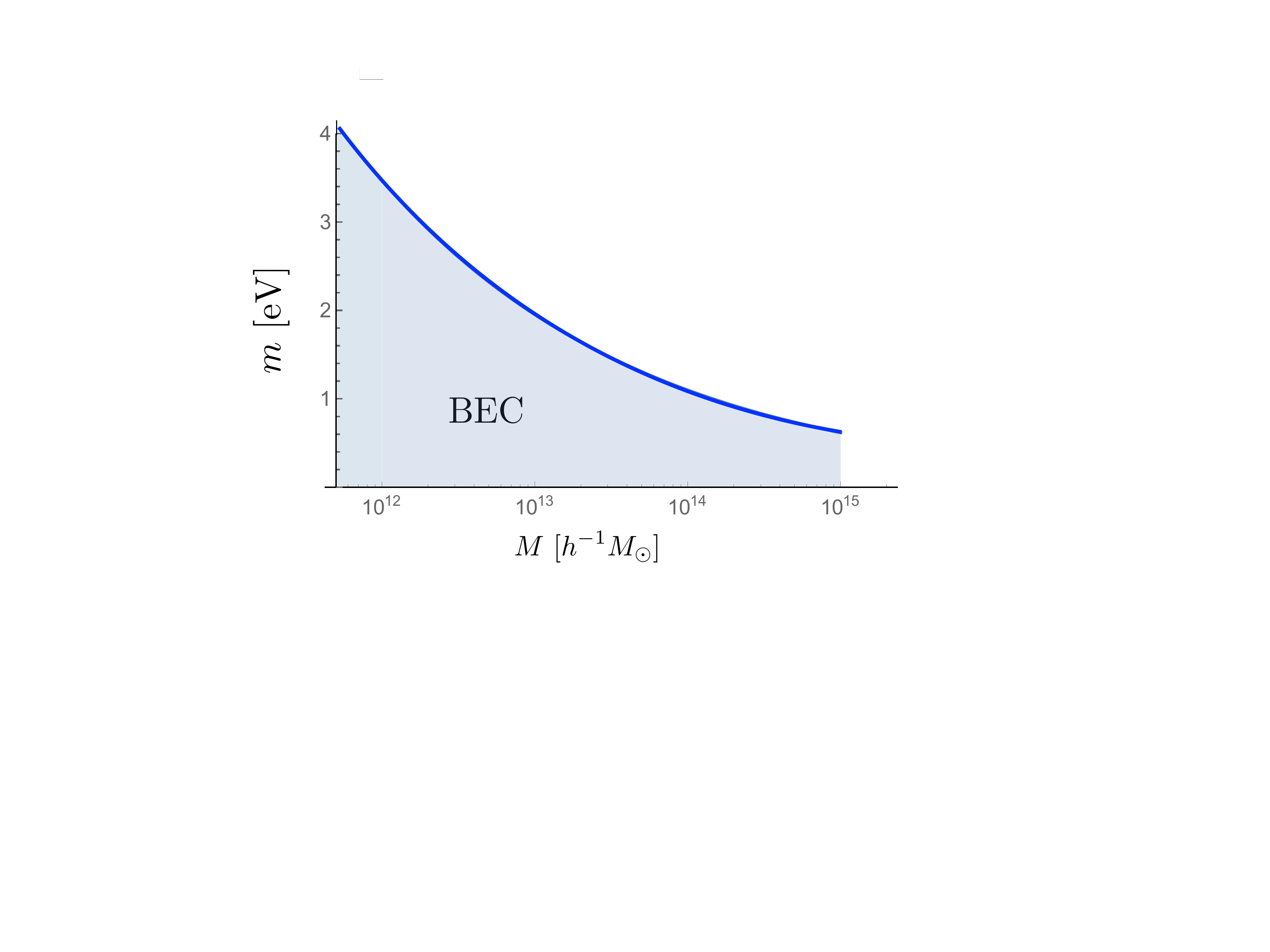}
\caption{\label{massdep} \small The shaded region corresponds to halos of mass $M$ within which DM forms a superfluid. For a given value of $m$,
superfluidity (and along with it the MOND phenomenon) only occurs in sufficiently low-mass halos, where the DM temperature (set by the virial velocity) is
below critical. This plot, reproduced from~\cite{Berezhiani:2015bqa}, assumes virialization at $z_{\rm vir} = 2$ and treats DM as a free Bose gas.}
\end{figure}

For starters, as mentioned in the Introduction the mass must be small enough such that DM forms a superfluid in galaxies~\cite{Berezhiani:2015pia,Berezhiani:2015bqa}:
\be
m \;\lsim\; 2~{\rm eV}\,.
\ee
In particular, our DM is comprised of axion-like particles. For a given value of $m$, superfluidity (and along with it MOND) only occurs in sufficiently low-mass halos, where the DM temperature (set by the virial velocity) is below critical. This is shown in Fig.~\ref{massdep}, reproduced from~\cite{Berezhiani:2015bqa}. The shaded region corresponds to halos that are sufficiently light (as a function $m$) to allow DM superfluidity. This plot assumes that halos virialize at $z_{\rm vir} = 2$ for concreteness. It also treats DM particles as a free Bose gas. A more refined calculation, including the gravitational potential
of the halo and contact interactions between DM particles, will be presented elsewhere~\cite{Anufuture}. For concreteness, we henceforth assume the fiducial value
\be
m = {\rm eV}\,. 
\ee
In that case massive galaxy clusters (with $M \;\gsim\; 10^{14}~M_\odot$) are above critical temperature, DM is in the normal phase, and there is no MOND phenomenon in clusters. 

\subsection{Adiabaticity and quantum corrections}

The wavefunction renormalization parameter $Z$ multiplying the kinetic term for $\chi$ is bounded from above and below by two considerations.
The first consideration is one of {\it adiabaticity}. The symmetry-breaking VEV~\eqref{broken} for $\chi$, which is key in deriving the MOND effective action, is exact if $|\vec{\nabla}\Phi|$ 
is constant. In realistic situations where $\Phi$ varies spatially,~\eqref{broken} will remain approximately true if the mass of $\chi$ around the minimum is large enough that $\chi$ can adiabatically adjust to changes in $|\vec{\nabla}\Phi|$. 

The physical mass around the symmetry-breaking minimum, $m_{\rm phys} \sim \frac{(\vec{\nabla}\Phi)^2M_{\rm Pl}}{Za_0}$, must be compared to $|\vec{\nabla}\ln \chi|$,
resulting in the adiabatic condition
\be
|\vec{\nabla}\ln \chi| \ll  \frac{(\vec{\nabla}\Phi)^2M_{\rm Pl}}{Za_0} \,.
\label{Zupper1}
\ee
Deep in the MOND regime and with spherical symmetry, we have $ |\vec{\nabla}\ln \chi|  = \frac{1}{r}$ and $(\vec{\nabla}\Phi)^2= a_{\rm N} a_0$,
hence~\eqref{Zupper1} implies
\be
Z \ll  a_{\rm N} M_{\rm Pl} r \,.
\ee
Rotation curves are measured down to $a_{\rm N} \sim 10^{-2}\;a_0$, see {\it e.g.},~\cite{McGaugh:2004aw}. Substituting the best-fit value~\eqref{a0} for $a_0$ and $r = {\rm kpc}$
for concreteness, we obtain
\be
Z \ll 10^{8}~{\rm GeV}\,.
\label{upper2}
\ee

The second consideration is one of {\it weak coupling}, both around the symmetry-restoring and symmetry-breaking vacua. As we will see, this condition is strongest around
$\chi = 0$. From the effective potential~\eqref{Veff18} we can read off the quartic coupling around $\chi = 0$ for the canonical field $\chi_c = Z \chi$:
\be
\lambda_{\rm Newton} \sim \frac{(\vec{\nabla}\Phi)^2}{Z^4M_{\rm Pl}^2} = \frac{a_{\rm N}^2}{Z^4M_{\rm Pl}^2}\,.
\ee
The last step uses the Newtonian result $|\vec{\nabla}\Phi | = a_{\rm N}$, which is consistent since 
we are working around $\chi = 0$. To be perturbative, this coupling should of course be small,
which translates to:
\be
Z\gg  \sqrt{a_{\rm N}M_{\rm Pl}}\,.
\label{lower1}
\ee
This is most stringent when $a_{\rm N}$ is large. Galactic observations of rotation curves and velocity dispersions extend to $a_{\rm N} \sim 10^2a_0 \sim 10^{-32}~{\rm eV}$~\cite{McGaugh:2004aw},
which gives
\be
Z \gg {\rm eV}\,.
\label{lower2}
\ee
Meanwhile, around the symmetry-breaking vacuum, the quartic coupling is
\be
\lambda_{\rm MOND} \sim \frac{|\vec{\nabla}\Phi|^5}{Z^4M_{\rm Pl}^2a_0^3} = \sqrt{\frac{a_{\rm N}}{a_0}} \frac{a_{\rm N}^2}{Z^4M_{\rm Pl}^2}\,.
\ee 
where we have used the MONDian form $|\vec{\nabla}\Phi | = \sqrt{a_{\rm N}a_0}$. Since $a_{\rm N} \ll a_0$ in the MOND regime, clearly we have $\lambda_{\rm MOND} \ll \lambda_{\rm Newton}$,
and hence~\eqref{lower2} gives the stronger bound, as claimed.

More generally, we should demand that quantum corrections to the $\chi$ effective potential are smaller than the tree-level potential~\eqref{Veff18},
such that the classical treatment is justified. The 1~loop effective potential receives the well-known Coleman-Weinberg
correction, $\Delta V_{1-{\rm loop}} \sim \frac{V_{,\chi\chi}^2}{Z^4}$, ignoring a logarithm factor. An explicit computation gives
\be
\frac{\Delta V_{1-{\rm loop}}}{V} \sim \lambda_{\rm Newton}  \frac{\left( \frac{3\chi^2 -1}{(1+\chi^2)^3} + \frac{(\vec{\nabla}\Phi)^2}{9a_0^2}\right)^2}{\frac{1}{1 + \chi^2} +  \frac{(\vec{\nabla}\Phi)^2}{9a_0^2}\chi^2}\,.
\label{V1loop}
\ee
The function multiplying $\lambda_{\rm Newton}$ depends on the value of $\frac{(\vec{\nabla}\Phi)^2}{a_0^2}$. In the MOND regime, $(\vec{\nabla}\Phi)^2 \ll a_0^2$, it is easily seen that $\frac{\Delta V_{1-{\rm loop}}}{V} \;\lsim\; \lambda_{\rm Newton}$, hence the condition~\eqref{lower1} enforcing $\lambda_{\rm Newton}\ll 1$ also ensures that quantum corrections are under control. In the
Newtonian regime, $(\vec{\nabla}\Phi)^2 \gg a_0^2$, however, the ratio~\eqref{V1loop} is largest near $\chi = 0$, where 
\be
\frac{\Delta V_{1-{\rm loop}}}{V} \sim \lambda_{\rm Newton} \frac{(\vec{\nabla}\Phi)^4}{a_0^4} \sim \frac{M_{\rm Pl}^2(\vec{\nabla}\Phi)^6}{Z^4a_0^4} \,.
\ee
This can be understood physically as correcting the Einstein-Hilbert term. Demanding that it be small imposes a lower bound on $Z$:
\be
Z \gg \sqrt{a_{\rm N}M_{\rm Pl}} \frac{a_{\rm N}}{a_0}\,,
\ee
where we have used the Newtonian result $|\vec{\nabla}\Phi| = a_{\rm N}$. This bound is tighter than~\eqref{lower1} by a factor of $\frac{a_{\rm N}}{a_0}\gg 1$.
Using $a_{\rm N} \sim 10^2a_0$ as a conservative value we obtain
\be
Z \gg 10^2~{\rm eV}\,.
\label{lower3}
\ee

Combining~\eqref{upper2} and~\eqref{lower3}, we see that the wavefunction renormalization must fall within the range
\be
10^2~{\rm eV} \ll Z \ll 10^{8}~{\rm GeV}\,.
\label{Zrange}
\ee
This results depends on coarse assumptions made about the range of $a_{\rm N}$ probed by galactic data and should be refined accordingly.
However there is clearly a broad range of allowed values that simultaneously satisfy the adiabaticity and weak coupling criteria.
Conversely, for a given value of $Z$ the adiabatic approximation will break down at sufficiently small acceleration. In particular if $Z$ lies near the upper end of~\eqref{Zrange}, the theory predicts departures from MOND in low-mass dwarfs, as suggested by observations~\cite{Spergel,Milgrom:1995hz,Angus:2008vs,Hernandez:2009by,Lughausen:2014hxa}. On the other hand, the application of MOND to these low-mass systems is complicated by astrophysical effects~\cite{Serra:2009tj,McGaugh:2010yr}. 

\subsection{Scale of the superfluid}

The last parameter is $\Lambda$, which sets the superfluid equation of state via~\eqref{eos}. The main requirement is that 
the superfluid pressure be sufficiently small to act as dust in all relevant situations. The ratio of pressure over energy density is
\be
w \equiv \frac{P}{\rho} =  \frac{1}{n}\frac{X}{m} =  \frac{1}{n} \left(\frac{\rho}{\Lambda^4}\right)^{\frac{1}{n-1}}\,.
\label{w}
\ee
The adiabatic sound speed of linear fluctuations is $c_s^2 = \frac{n}{n-1}w$. At low density ($\rho \ll \Lambda^4$) the superfluid behaves as dust,
whereas at high density ($\rho \gg \Lambda^4$) it behaves as a relativistic component. (Of course~\eqref{w} ceases to be valid in the relativistic regime.)

At the very least we should impose that the equation of state is small by the time of matter-radiation equality, $w_{\rm eq} \ll 1$, when the density is $\rho = \rho_{\rm eq} \simeq 0.4~{\rm eV}^4$:
\be
w_{\rm eq} =  \frac{(0.4)^{\frac{1}{n-1}}}{n} \left(\frac{\Lambda}{{\rm eV}}\right)^{-\frac{4}{n-1}} \ll 1\,.
\label{weqbound}
\ee
Subsequently $\rho$ redshifts as $1/a^3$, hence $w$ decreases in time as 
\be
w \sim \frac{1}{a^{3(n-1)}}\,. 
\label{wevolve}
\ee
Therefore if $w$ is small at equality, it will become even smaller afterwards. Furthermore, the DM density in galaxies is smaller than $ \rho_{\rm eq}$, hence if DM behaves as dust by equality
it will also behave as dust in galaxies, as desired. Therefore~\eqref{weqbound} is a necessary and sufficient condition on the equation of state. 

A precise upper bound on $w_{\rm eq}$ from cosmological data, such as the cosmic microwave background, would require a detailed analysis. For concreteness,
we will assume $w_{\rm eq} \;\lsim\; 10^{-2}$ is sufficient. In that case,~\eqref{weqbound} implies the following bound on $\Lambda$ (assuming $n$ of order unity):
\be
\Lambda \;\gsim\; 2 \times 10^{\frac{n-2}{2}}~{\rm eV}\,.
\label{Lambdalimit}
\ee
In other words, with $\Lambda \;\gsim \; {\rm eV}$ our superfluid DM behaves as dust sufficently early in the history of the universe. We will have more to say about cosmology in Sec.~\ref{cosmo}. 
In the next Section we will study the superfluid density profile in galaxies and discover that $\Lambda \sim {\rm eV}$ results
in halos of realistic size for the fiducial values $n = 2$ and $5/2$. It is quite remarkable that with $m$ and $\Lambda$ both of order ${\rm eV}$ it is possible
to simultaneously obtain an acceptable cosmology and realistic-size halos. Incidentally, since $c_s \ll 1$ we see from~\eqref{Lamstrong} that the strong coupling scale
$\Lambda_{\rm strong}$ is $\ll {\rm eV}$, hence as a function of increasing energy the superfluid effective theory breaks down before
the superfluid becomes relativistic.

\section{MONDian Density Profile}
\label{profile}

In this Section we calculate the DM superfluid density profile for a static, spherically-symmetric halo, assuming hydrostatic static equilibrium. 
This is DM-only calculation, ignoring baryons. A key difference from BK is that the
effective modification to gravity is now universal, affecting not only baryons but also DM. Another difference is that the superfluid equation of state
is more general, as mentioned earlier.

\begin{figure}[t]
\centering
\includegraphics[width=4.5in]{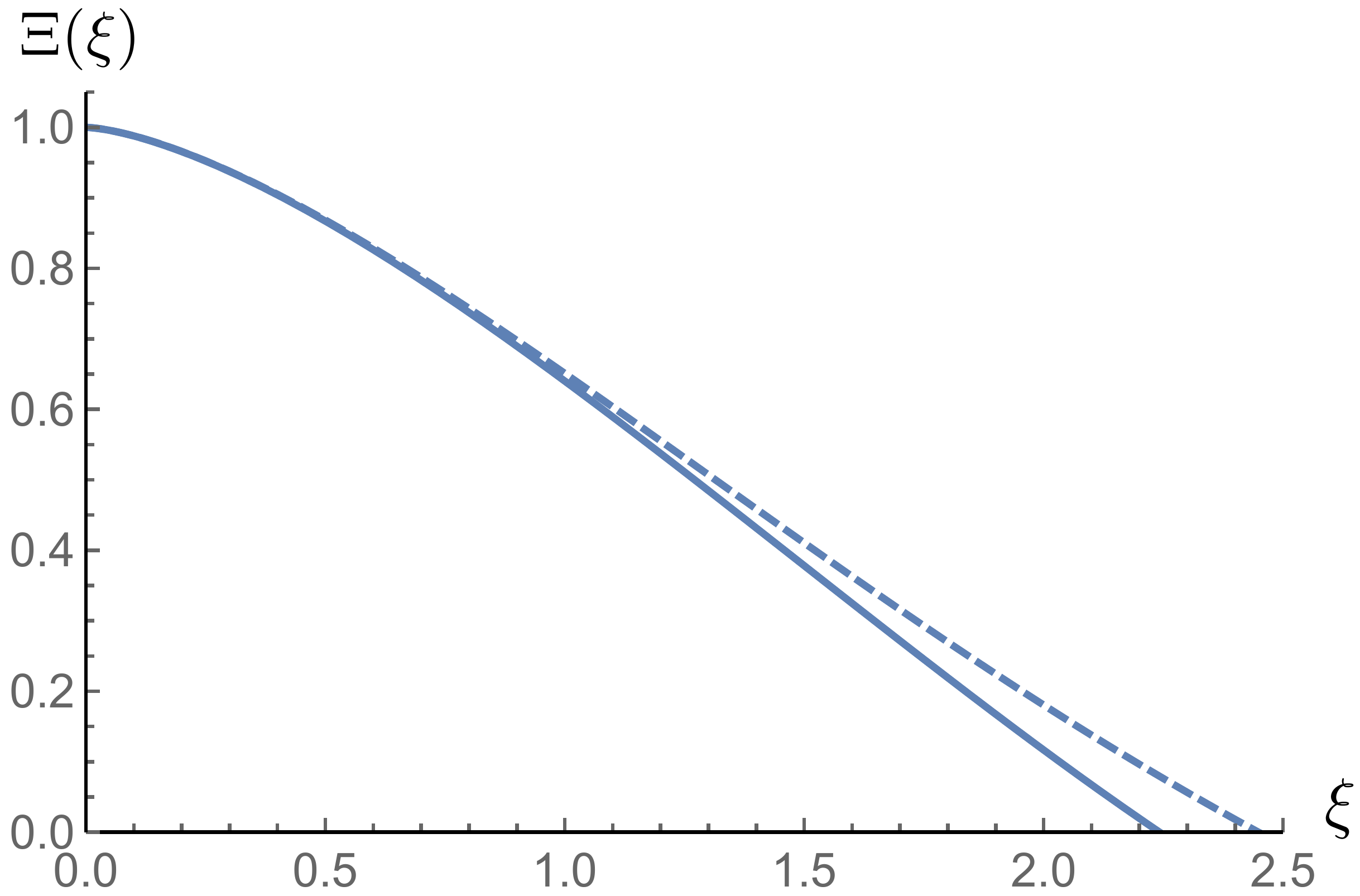}
\caption{\label{MONDlaneemden} \small Numerical solution to the hydrostatic equation~\eqref{nonlinLaneEmdenEqn}, which is a MONDian generalization of the Lane-Emden equation,
for $n=2$ (solid) and $n=5/2$ (dashed). The boundary conditions are $\Xi(0) = 1$ and $\Xi'(0) = 0$. The value $\xi_1$ at which each solution vanishes, together with the first derivative ${\rm d}\Xi/{\rm d}\xi_1$ at that point, are listed in Table~\ref{xivalues}.}
\end{figure}

For simplicity, we derive the DM profile for halos that are sufficiently light that the MOND law $a = \sqrt{a_{\rm N}a_0}$ applies
throughout. At the end of the calculation we will determine the mass range over which this
approximation is valid. The equation of hydrostatic equilibrium is then
\be
\frac{P'(r)}{\rho(r)} = -\sqrt{\frac{4\pi G_{\rm N}a_0}{r^2} \int_0^r {\rm d}\tilde{r} \,\tilde{r}^2\rho(\tilde{r})}\,,
\label{hydro}
\ee
where $'\equiv \frac{{\rm d}}{{\rm d}r}$. Recall that $\rho = m P_{,X}$, hence the left-hand side is simply $\frac{P'}{\rho} = X'(r)$ for any $P$.
On the right-hand side, focusing on the power-law form~\eqref{Pn} the superfluid density is given by~\eqref{rhosmu}: $\rho = \Lambda^4 \left(\frac{X}{m}\right)^{n-1}$.
Squaring both sides,~\eqref{hydro} can be expressed as 
\be
r^2X'^2 = \frac{4\pi G_{\rm N} a_0 \Lambda^4}{m^{n-3}}  \int_0^r {\rm d}\tilde{r}\, \tilde{r}^2X^{n-1}(\tilde{r})\,.
\label{hydro2}
\ee
This in turn implies the differential equation
\be
\frac{1}{r^2} \left(r^2 X'^2\right)' = \frac{4\pi G_{\rm N} a_0 \Lambda^4}{m^{n-3}}X^{n-1}\,.
\ee
In terms of dimensionless variables
\be
\Xi  \equiv \frac{X}{X_0}\,;\qquad \xi \equiv  \left(4\pi G_{\rm N}a_0\Lambda^4\right)^{1/3} X_0^{\frac{n-3}{3}}~r\,,
\label{dimless}
\ee
with $X_0$ denoting the value at the origin, the differential equation becomes
\be
\frac{1}{\xi^2}\frac{{\rm d}}{{\rm d}\xi} \left(\xi^2 \left(\frac{{\rm d}\Xi}{{\rm d}\xi}\right)^2\right) = \Xi^{n-1}\,.
\label{nonlinLaneEmdenEqn}
\ee
This can be thought of as a MONDian generalization of the Lane-Emden equation, $\frac{1}{\xi^2}\frac{{\rm d}}{{\rm d}\xi} \left(\xi^2 \frac{{\rm d}\Xi}{{\rm d}\xi}\right) = - \Xi^{n}$,
which is the equation obtained assuming Newtonian gravity. 

Equation~\eqref{nonlinLaneEmdenEqn} can be integrated similarly to the standard Lane-Emden equation. The boundary conditions are $\Xi(0) = 1$ (by definition) and $\Xi'(0) = 0$ (to ensure smoothness at the origin). The numerical solution is shown in Fig.~\ref{MONDlaneemden} for $n = 2$ and~$5/2$. The density profile is cored, by virtue of the
boundary conditions, and has finite extent. From the numerical solution we can read off the value $\xi_1$ at which the density vanishes, together with the first derivative
at that point (which will come in handy shortly). These are listed in Table~\ref{xivalues}.

The value $\xi_1$ sets the physical size of the superfluid core through~\eqref{dimless}:
\be
R = \left(\frac{1}{4\pi G_{\rm N}a_0\Lambda^4}\right)^{1/3}  \left(\frac{\rho_0}{\Lambda^4}\right)^{-\frac{n-3}{3(n-1)}}~\xi_1\,,
\label{Rdef}
\ee
where we have substituted~\eqref{rhosmu}. It is useful to express the central density $\rho_0$ in terms of the halo mass $M$. Using~\eqref{hydro2}, the mass enclosed is
\be
M = \frac{r^2X'^2}{m^2G_{\rm N}a_0} = \frac{X^2_0}{m^2G_{\rm N}a_0}\xi^2_1 \left(\frac{{\rm d}\Xi}{{\rm d}\xi_1}\right)^2\,.
\ee
Inverting this to find $X_0$, and hence $\rho_0$, as a function $M$ gives
\be
\frac{\rho_0}{\Lambda^4}  \simeq  \left(\frac{M}{10^{12}\,M_\odot}\right)^{\frac{n-1}{2}} \left(\xi_1\left\vert\frac{{\rm d}\Xi}{{\rm d}\xi_1}\right\vert\right)^{-(n-1)} 10^{-6(n-1)}\,,
\ee
where we have substituted the best-fit critical acceleration~\eqref{a0}. Plugging this back into~\eqref{Rdef} gives $R$ as a function of $M$. Using the cosmological
constraint~\eqref{Lambdalimit} on $\Lambda$, we obtain an upper bound on the halo radius:
\be
R \;\lsim\;  \left(\frac{M}{10^{12}\,M_\odot}\right)^{\frac{3-n}{6}}  \xi_1^{n/3}\left\vert\frac{{\rm d}\Xi}{{\rm d}\xi_1}\right\vert^{\frac{n-3}{3}} 10^{\frac{4n-5}{3}}\;{\rm kpc}\,.
\label{haloRbound}
\ee

\begin{table}
\centerline{
\small
\begin{tabular}{| l | c | c | }\hline 
\raisebox{8pt} {\phantom{M}}~~~~~~~\raisebox{-8pt}{\phantom{M}} & \raisebox{8pt} {\phantom{M}}$\xi_1$  \raisebox{-8pt}{\phantom{M}}  & \raisebox{8pt} {\phantom{M}} $ \left\vert\frac{{\rm d}\Xi}{{\rm d}\xi_1}\right\vert$   \raisebox{-8pt}{\phantom{M}} \\ \hline
\raisebox{8pt} {\phantom{M}}	$n=2$  \raisebox{-8pt}{\phantom{M}} & \raisebox{8pt} {\phantom{M}}$2.25$  \raisebox{-8pt}{\phantom{M}} & \raisebox{8pt} {\phantom{M}}$0.46$  \raisebox{-8pt}{\phantom{M}}  \\				\hline
\raisebox{8pt} {\phantom{M}}	$n=5/2$ \raisebox{-8pt}{\phantom{M}} & \raisebox{8pt} {\phantom{M}} $2.45$  \raisebox{-8pt}{\phantom{M}} &\raisebox{8pt} {\phantom{M}} $0.37$  \raisebox{-8pt}{\phantom{M}} \\ 							\hline
\end{tabular}
}
\caption{\small Values of the radius $\xi_1$ at which the density profile vanishes, together with the first derivative of the density at that point, for our fiducial $n$ values.}
\label{xivalues}
\end{table}

Let us study our fiducial values of $n$ more closely:

\begin{itemize}

\item {\bf Standard BEC case ($n = 2$):} In this case~\eqref{haloRbound} gives
\be
R \;\lsim\; \left(\frac{M}{10^{12}\,M_\odot}\right)^{1/6}~20~{\rm kpc} \,.
\label{R2}
\ee
This is marginally acceptable for fitting rotation curves with MOND.\\

\item {\bf UFG case ($n = 5/2$):} In this case we have
\be
R \;\lsim \;  \left(\frac{M}{10^{12}\,M_\odot}\right)^{1/12}~115~{\rm kpc} \,.
\label{R52}
\ee
This is definitely sufficient for fitting rotation curves. 

\end{itemize}

For consistency, we should check that the halos considered are indeed in the MONDian regime, $a_{\rm N} < a_0$,
throughout their volume. Since the density profile is nearly constant, this condition is most stringent at the edge of the halo, that is,
\be
\frac{a_{\rm N}}{a_0} = \frac{G_{\rm N} M}{a_0R^2} < 1\,.
\ee
Substituting~\eqref{haloRbound} and ignoring factors of order unity, this translates to an upper bound on the halo mass:
\be
M \;\lsim\; \xi_1^{2}\left\vert\frac{{\rm d}\Xi}{{\rm d}\xi_1}\right\vert^{\frac{2(n-3)}{n}}10^{3(n+1)}\,M_\odot \,.
\ee
Using Table~\ref{xivalues} we find $M \;\lsim\; 10^{10}~M_\odot$ for $n =2$ and $M \;\lsim\; 10^{11}~M_\odot$ for $n =5/2$. Combined with the above discussion, we see that 
$n=2$ and $n=5/2$ are phenomenologically suitable to achieve the MOND phenomenon in a broad enough range of galactic masses and sizes. It is interesting to note that 
the mass dependence is quite weak, particularly for $n=5/2$, hence halos have nearly universal size .

We close with a remark on the Baryonic Tully-Fisher Relation. Recall that this relates the total {\it baryonic} mass to the asymptotic circular
velocity as $v^4_{\rm c}\sim M_{\rm b}$. In ``standard MOND", there is of course no DM, and this relation follows as an exact prediction. In our case,
on the other hand, both DM and baryons experience MOND, hence the predicted relation involves the total mass, $v^4_{\rm c}\sim M_{\rm tot}$. This
could run afoul of observations unless DM is subdominant in the region where $v_{\rm c}$ is measured. Although this likely requires a careful analysis,
two factors work in our favor. First, precisely because DM is subject to MOND, the actual DM fraction is less than the inferred fraction assuming Newtonian
dynamics, by the usual factor:
\be
M_{\rm inferred} = M \sqrt{\frac{a_0}{a_{\rm N}}}\,.
\label{Minferred}
\ee
Second, as we have seen the density profile is nearly constant in the inner region of the halo where rotation curves are probed. Hence the mass
enclosed scales as $M(r) \sim r^3$ and therefore decreases rapidly as $r\rightarrow 0$. 

For concreteness, consider a large disk galaxy with $M_{\rm b} = 10^{10}~M_\odot$ whose rotation curve is measured out to $r_{\rm max} = 20~{\rm kpc}$. We focus on the case $n =5/2$, though this is 
admittedly the more favorable of the two cases. We assume a comparable superfluid DM mass of $M = 10^{10}~M_\odot$ and a superfluid core radius of $R = 100$~kpc, consistent with~\eqref{R52}. From~\eqref{Minferred} the inferred mass at the superfluid radius (say from lensing observations) is $M_{\rm inferred} \sim 3\times 10^{11}~M_\odot$, corresponding to an effective mass-to-light ratio of 30. 
Within the maximal radius probed by rotation curves, however, the enclosed DM mass is suppressed by a factor of $\left(\frac{r_{\rm max}}{R}\right)^3 \simeq 10^{-2}$ relative to $M_{\rm b}$, and
the Tully-Fisher relation holds to good accuracy.

\section{Relativistic Gravitational Theory and Lensing}
\label{lensing}

In this Section we describe a possible covariantization of the theory~\eqref{Ltot}. An important constraint is to preserve the equality of gravitational potentials, $\Phi = \Psi$, in order for dynamical and lensing mass estimates to coincide. In TeVeS~\cite{Bekenstein:2004ne} and other relativistic completions of MOND, this requires introducing a unit time-like vector field $A_\mu$, together with fairly intricate couplings between ordinary matter and the scalar, vector and tensor fields. In BK, the story is somewhat simpler. The 4-velocity $u^\mu$ of the normal DM component (relative to the superfluid component)
offers a unit time-like vector. There is no need to postulate an additional vector field. However, the form of the coupling to matter remains complicated.
In our case, we will also make use of $u^\mu$, but the necessary complications will occur in the gravitational action (since we are effectively working in Jordan frame).

Let us start with the LO action~\eqref{Pn} and see how it arises as the non-relativistic limit of a Lorentz invariant superfluid. 
At low energy a relativistic superfluid is described by the Goldstone field $\Theta$, related to the non-relativistic
phase $\theta$ by
\be
\Theta = mt + \theta\,.
\ee
Its effective action to lowest order in derivatives is a function of the Lorentz
invariant kinetic term ${\cal Y} \equiv - \frac{1}{2} (\partial\Theta)^2$. In the
non-relativistic limit, this can be expanded as
\be
{\cal Y} = - \frac{1}{2} (\partial\Theta)^2 \simeq \frac{m^2}{2} + m X\,,
\ee
where, as before, $X =  \dot{\theta}  - m \Phi - \frac{(\vec{\nabla}\theta)^2}{2m}$.
Therefore the LO action~\eqref{Pn} arises as the non-relativistic limit of
\be
{\cal L}_{\rm LO} = \frac{\Lambda^4}{n} \left(\frac{{\cal Y}}{m^2} - \frac{1}{2}\right)^n\,.
\label{Pnrel}
\ee

Similarly, it is easy to see that the NLO operator $(\vec{\nabla}X)^2$ descends from
\be
(\partial {\cal Y})^2 \simeq m^2 (\vec{\nabla}X)^2\,.
\ee
Consider the gravitational mixing terms in the NLO action~\eqref{NLOX}:
\be
{\cal L}_{\rm grav}  = -M_{\rm Pl}^2(\vec{\nabla} \Phi)^2 F\left(\chi,(\vec{\nabla}X)^2\right)\,;\qquad F\left(\chi,(\vec{\nabla}X)^2\right) = \frac{1}{1 + \chi^2} + \frac{(\vec{\nabla} X)^2}{9m^2a_0^2}\chi^2\,.
\label{Lgravgen}
\ee
An obvious guess for its relativistic generalization is 
\be
{\cal L}_{\rm naive}  = \frac{M_{\rm Pl}^2}{2}R \,F\left(\chi,\frac{(\partial {\cal Y})^2}{m^2}\right)\,.
\label{Lnaive}
\ee
However this naive choice is inconsistent with the lensing constraint $\Phi = \Psi$. To see this, substitute the weak-field static metric ${\rm d}s^2 = -(1+ 2\Phi) {\rm d}t^2 + (1-2\Psi){\rm d}\vec{x}^2$
and expand the action to quadratic order. After integration by parts (taking into account the measure $\sqrt{-g}$), the result is
\be
{\cal L}_{\rm naive} \simeq M_{\rm Pl}^2\Bigg\{\bigg( \left(\vec{\nabla} (\Psi -\Phi)\right)^2 - (\vec{\nabla} \Phi)^2\bigg)\,F + \vec{\nabla} F \cdot \vec{\nabla} \Phi-2\vec{\nabla} F \cdot  \vec{\nabla} \Psi \Bigg\}\,.
\label{Lnaivequad}
\ee
Varying with respect to $\Psi$ gives a constraint, whose solution is $\Psi = \Phi + \ln F$. In other words, because of the non-minimal coupling, we find $\Psi \neq \Phi$
(unless of course $F = {\rm const}$, in which case~\eqref{Lnaive} is just General Relativity).\footnote{Mapping~\eqref{Lnaive} to Einstein frame,
$g^{\rm E}_{\mu\nu} = F g_{\mu\nu}$, the gravitational potentials are related by $\Phi_{\rm E} = \Phi + \frac{1}{2} \ln F$,
$\Psi_{\rm E} = \Psi - \frac{1}{2} \ln F$. In Einstein frame, we of course have $\Psi_{\rm E} = \Phi_{\rm E}$, which is consistent with the result $\Psi = \Phi + \ln F$.}

It is evident that the problem stems from the last two terms~\eqref{Lnaivequad}.
If these terms could be somehow eliminated, such that the action reduced to 
\be
{\cal L}_{\rm desired} \simeq M_{\rm Pl}^2\bigg( \left(\vec{\nabla} (\Psi -\Phi)\right)^2 - (\vec{\nabla} \Phi)^2\bigg)\,F \,,
\label{Ldesired}
\ee
then varying with respect to $\Psi$ would give $\Psi = \Phi$. And substituting this back into~\eqref{Ldesired} would yield~\eqref{Lnaive}, as desired.

What covariant action can possibly give rise to~\eqref{Ldesired} in the weak-field limit? There are likely many possibilities, but as a proof of principle
the following construction does the job. The construction makes use of the normal DM component, which, albeit subdominant to the superfluid component,
is expected to be present at some level in actual galactic halos, given that DM has non-zero temperature. At sub-critical temperature
the system is described phenomenologically by Landau's two-fluid model: an admixture of a superfluid component, which has zero viscosity, and a
normal component, which is viscous and carries entropy. Their relative fraction is a function of
temperature, and hence the mass of the collapsed object.

For simplicity, we model the normal component as a perfect fluid. From a field theory standpoint, it is described by 3 Lorentz scalars $\phi^I(x^\mu)$, $I = 1,2,3$,
specifying the comoving position of each fluid element as a function of ``laboratory" space-time coordinates $x^\mu$. The ground state configuration is $\phi^I = x^I$,
with small fluctuations $\pi^I = \phi^I - x^I$ describing phonons. By virtue of being perfect, the effective theory for the $\phi^I$'s should
be invariant under arbitrary shear deformations: $\phi^I \rightarrow \hat{\phi}^I$, with $\det \frac{\partial \hat{\phi}^I}{\partial \phi^J} = 1$. At lowest number in derivatives,
this means that the action can only depend on the 3-density:
\be
n \equiv \sqrt{\det \left(g^{\mu\nu}\partial_\mu\phi^I \partial_\nu\phi^J\right)}\,.
\ee
In the fluid ground state and in the weak-field quasi-static approximation, this reduces to
\be
n \simeq  1 + 3\Psi\,.
\ee
In particular, the operator $\partial_\mu F \partial^\mu n \simeq 3 \vec{\nabla}F \cdot \vec{\nabla}\Psi$
can be used to cancel the last term in~\eqref{Lnaivequad}. The next-to-last term, meanwhile, can be canceled by the superfluid
term $\partial_\mu F \frac{\partial^\mu{\cal Y}}{m^2}\simeq   \frac{\vec{\nabla}X}{m}\cdot \vec{\nabla} F$. 
Therefore, an action that covariantizes~\eqref{Lgravgen} and accomplishes what we want is (after integration by parts):
\be
{\cal L}_{\rm grav}  =M_{\rm Pl}^2\left\{ \frac{R}{2} - \left(\frac{2}{3}\Box n + \frac{\Box {\cal Y}}{m^2}\right)\right\}F \,.
\label{Lgravcov}
\ee
Of course this action leads to equations of motions that are higher than second-order in time, but by construction the mass of the would-be ghost must diverge as $c_s\rightarrow 0$,
and hence lies above the cutoff of the effective theory.  
 
A fully relativistic generalization of~\eqref{Ltot} is given by
\bea
\nonumber
{\cal L} & = &M_{\rm Pl}^2\left\{ \frac{R}{2} - \left(\frac{2}{3}\Box n + \frac{\Box {\cal Y}}{m^2}\right)\right\}F - \frac{1}{2} Z^2 (\partial \chi)^2 +\frac{\Lambda^4}{n} \left(\frac{{\cal Y}}{m^2} - \frac{1}{2}\right)^n + {\cal L}_{\rm b} \,;\\
F &\equiv &  \frac{1}{1 + \chi^2} + \frac{(\partial {\cal Y})^2}{9m^4a_0^2}\chi^2\,,
\label{Ltotfinal}
\eea
where ${\cal L}_{\rm b}$ is the action for ordinary matter (baryons). In the weak-field quasi-static limit, this action implies the equality of gravitational and lensing potentials, and
reduces to~\eqref{Ltot} in the non-relativistic regime.  

\section{Cosmology}
\label{cosmo}

In this Section we discuss the cosmological implications of DM/MOND superfluidity. Let us begin with a few general remarks that carry over from the BK scenario. 
Our DM component consists of axion-like particles with $m \sim {\rm eV}$, hence they must be produced out of equilibrium and remain decoupled from ordinary matter
throughout the cosmological history. Thanks to its self-interactions, the DM component rapidly reaches thermal equilibrium with itself, at a temperature far smaller
than that of the baryon-photon fluid, and becomes a superfluid. Cosmologically it remains superfluid forever after. 

A natural genesis mechanism is the so-called vacuum displacement mechanism familiar from axion physics. In the early universe when $H \gg m$, the $U(1)$
scalar field is initially displaced from its minimum and is overdamped by Hubble friction. Once $H$ drops to a value of order $m$, the field starts oscillating,
thereby converting potential energy into DM particles. As pointed out by BK, with $m\sim {\rm eV}$ the DM genesis interestingly occurs at a baryon temperature
of order the weak scale, $T_{\rm i}^{\rm baryons} \sim \sqrt{m M_{\rm Pl}} \sim 50$~TeV. 

Next let us study the behavior of the superfluid component on a cosmological background. For the moment we will ignore the mixing with $\chi$
and describe the superfluid using the LO action~\eqref{Pn}. (We will justify this momentarily.) Varying with respect to $\theta$ gives
the equation of motion
\be
\frac{{\rm d}}{{\rm d}t} \left(a^3\dot{\theta}^{n-1}\right) = 0\,,
\label{thetaeomcosmo}
\ee
hence $\dot{\theta} = X \sim 1/a^{3(n-1)}$. The superfluid density~\eqref{rhosmu} scales as dust, $\rho \sim 1/a^3$, as it should.
Of course the non-relativistic approximation implicit in~\eqref{thetaeomcosmo} is only valid provided $X \ll m$, which corresponds to
$w \ll 1$. So far, so good.

At NLO the key question is what happens to the symmetron field $\chi$? In particular, is the universe on cosmological
scales in the Einstein gravity, symmetry-restoring phase ($\chi = 0$), or in the MONDian gravity, symmetry-broken
phase? To answer this, consider the effective potential which can be read off from~\eqref{Ltotfinal}:
\be
V (\chi)  = - M_{\rm Pl}^2\left\{ \frac{R}{2} - \left(\frac{2}{3}\Box n + \frac{\Box {\cal Y}}{m^2}\right)\right\}\,F\,.
\ee
On a cosmological background and in the normal fluid ground state, we have $n \sim 1/a^3$,
hence $\Box n = 0$. Meanwhile, since $\dot{\cal Y} = m\dot{X} \sim H\dot{X}$ the last term gives 
\be
\frac{\Box {\cal Y}}{m^2} \sim H^2 \frac{X}{m} \ll H^2\,,
\ee
and is therefore negligible compared to $R \sim H^2$. Hence, during matter domination the effective potential reduces to
\be
V (\chi)  \simeq  - \frac{M_{\rm Pl}^2}{2} R\, F =  - \frac{3}{2}H^2M_{\rm Pl}^2\,F\,.
\ee
In particular, using the explicit form for $F$ the mass around $\chi = 0$ is
\be
m_\chi^2 = 3H^2M_{\rm Pl}^2 \left(1 + \frac{\dot{\cal Y}^2}{9m^4a_0^2}\right) \,.
\ee
The mass squared is positive definite. {\it Therefore, cosmologically the universe is always in the Einstein-gravity, symmetry-restoring phase.} The non-minimal term $\sim \frac{R}{1+\chi^2}$,
while having a destabilizing effect $\chi$ in the static regime (c.f., first term in~\eqref{meffstatic}), flips sign in the time-dependent, cosmological regime
and thus has a stabilizing effect. 

Consequently, the background expansion history
and linear growth of density perturbations are indistinguishable from $\Lambda$CDM cosmology. This is an important difference from the BK scenario,
where, because of the direct phonon-baryon coupling, it was necessary to invoke temperature-dependent $a_0$ and
coupling parameter for consistency with cosmological observations~\cite{Berezhiani:2015pia,Berezhiani:2015bqa}. In the
present case the theory naturally reverts to Einstein gravity on cosmological scales, thanks to the symmetry-restoring
mechanism. 

\section{Conclusions}
\label{conclu}

In this paper we proposed an alternative mechanism for MOND to emerge from a phase of DM superfluidity. The mechanism has much in common 
with the original BK scenario: DM consists of strongly-interacting, axion-like particles, which form a superfluid in galaxies. However, instead of relying on the
mediation of a ``fifth force" by phonons, the mechanism described here uses NLO corrections in the effective theory of a superfluid to
generate a MOND-like action directly in terms of the gravitational potential. This allows more freedom in specifying the superfluid equation of state (encoded in the LO action),
including the well-known cases $n=2$ (standard superfluid) and $n = 5/2$ (UFG). Furthermore, because phonons need no longer couple directly with matter,
their action now enjoys an exact shift symmetry, as expected for the phase of a wavefunction. 

To be fair, at some level there is conservation of trouble --- the lack of freedom in specifying the LO action in BK has been traded for a lack of freedom in specifying the NLO terms. 
The truth is that {\it any} formulation of MOND requires a special form for the action. Usually this is in the guise of a special modification to the gravitational sector. Here
the required modification is in the effective theory of the DM superfluid. Be that as it may, we were able to argue that the effective theory is under control, in the sense that
quantum corrections are consistently small both in the Newtonian and deeply MONDian regimes. 

Phenomenologically the most important difference with BK is that the MOND force law applies to both baryons and DM. This affects the DM superfluid density profile,
which obeys a generalization of the Lane-Emden equation appropriate for MONDian gravity. 

Our explicit realization of MOND uses a symmetron-inspired mechanism, which we believe is novel. The action involves a $\mathbb{Z}_2$-invariant scalar field $\chi$,
coupling both non-minimally to gravity and to superfluid NLO corrections. The action is analytic in all field variables, with the eventual non-analyticity of 
the MOND effective action resulting from spontaneous symmetry breaking of the discrete $\mathbb{Z}_2$ symmetry. Specifically, for $|\vec{\nabla}\Phi| > 3a_0$
the mass squared is positive, the symmetry is unbroken, and the action reduces to Newtonian gravity. For $|\vec{\nabla}\Phi| < 3a_0$, the mass squared is negative, the symmetry is
spontaneously broken, and the action reduces to MONDian gravity. 

The above mechanism assumes a quasi-static regime, which is applicable in non-linear structures. On cosmological scales, however, 
time dependence is obviously important, and the mass of $\chi$ turns out to be positive definite. In other words, as far as the background expansion
and the linear growth of density perturbations are concerned, the universe is always in the Einstein-gravity, symmetry-restoring phase. The expansion and linear growth histories are indistinguishable from $\Lambda$CDM cosmology.

In on-going work~\cite{future} we are currently studying numerical simulations of DM superfluidity to explore its impact on non-linear structure formation.
There are a number of observational signatures (common to BK and the present model) that can potentially distinguish our scenario from ordinary MOND
and/or standard $\Lambda$CDM: numerous low-density vortices in galaxies; merger dynamics depending on the infall velocity {\it vs} phonon sound speed;
distinct mass peaks in bullet-like cluster mergers, corresponding to superfluid and normal components; interference patterns in sub-critical mergers, {\it etc.}

The elephant in the room is of course dark energy. Certainly it would be very compelling if cosmic acceleration could emerge from the same underlying substance as DM and MOND. Preliminary ideas along these lines will appear in a forthcoming publication~\cite{future2}.\\

\noindent {\bf Acknowledgements:} We thank Benoit Famaey, Tom Lubensky and Moti Milgrom for helpful discussions, and especially Lasha Berezhiani for numerous discussions
and initial collaboration. This work is supported in part by NSF CAREER Award PHY-1145525, NASA ATP grant NNX11AI95G and a New Initiative Research Grant from the
Charles E. Kaufman fund of The Pittsburgh Foundation.

\end{document}